\documentclass[aps,pra,superscriptaddress,twocolumn,floatfix,a4paper]{revtex4}

\usepackage{graphicx,graphics,epsfig}   
\usepackage{dcolumn}    
\usepackage{bm}         
\usepackage{amsmath}    
\usepackage{verbatim}   
\usepackage{color}      
\usepackage{subfigure}  
\usepackage{times,natbib}
\usepackage{amsmath,amsfonts,amssymb,graphics,graphics,color,times}

\usepackage{latexsym}
\usepackage{amsmath}
\usepackage{amssymb}
\usepackage{amsfonts}
\usepackage{amsthm}
\usepackage{mathrsfs}
\usepackage{color,verbatim,graphics}
\usepackage{psfrag}
\DeclareMathAlphabet{\mathrsfs}{U}{rsfs}{m}{n}
\DeclareMathAlphabet{\mathpzc}{OT1}{pzc}{m}{it}
\DeclareMathAlphabet{\matheus}{U}{eus}{m}{n}
\DeclareMathAlphabet{\mathbbold}{U}{bbold}{m}{n}

\def\one{\leavevmode\hbox{\small1\normalsize\kern-.33em1}}

\newcommand{\ba}{\begin{eqnarray}}
\newcommand{\ea}{\end{eqnarray}}
\newcommand{\ban}{\begin{eqnarray*}}
\newcommand{\ean}{\end{eqnarray*}}

\newcommand{\ket}[1]{|#1\rangle}
\newcommand{\bra}[1]{\langle#1|}

\newcommand{\ketbra}[2]{|#1\rangle\langle#2|}

\begin{document}

\title{A class of genuinely high-dimensionally entangled states with a positive partial transpose}

\author{K\'aroly F. P\'al}
\affiliation{Institute of Nuclear Research of the Hungarian
Academy of Sciences H-4001 Debrecen, P.O. Box 51, Hungary}

\author{Tam\'as V\'ertesi}
\affiliation{Institute of Nuclear Research of the Hungarian
Academy of Sciences H-4001 Debrecen, P.O. Box 51, Hungary}

\date{\today}


\begin{abstract}
Entangled states with a positive partial transpose (so-called PPT states) are central to many interesting problems in quantum theory. On one hand, they are considered to be weakly entangled, since no pure state entanglement can be distilled from them. On the other hand, it has been shown recently that some of these PPT states exhibit genuinely high-dimensional entanglement, i.e. they have a high Schmidt number. Here we investigate $d\times d$ dimensional PPT states for $d\ge 4$ discussed recently by Sindici and Piani, and by generalizing their methods to the calculation of Schmidt numbers we show that a linear $d/2$ scaling of its Schmidt number in the local dimension $d$ can be attained. 
\end{abstract}

\maketitle

\section{Introduction}
\label{Intro} 

Entanglement is at the heart of quantum theory and is also a key resource in quantum information applications~\cite{horo_review,GT_review}. Bipartite systems which are entangled across many degrees of freedom are especially interesting in this respect. Firstly, they are usually more robust to noise than systems with less degrees of freedom~\cite{grob,dada}. Secondly, they allow us to devise protocols which are genuinely high-dimensional. In particular, low dimensional systems are not enough for these protocols to work~\cite{lanyon}.

Experimentally, high-dimensional entanglement also became feasible in the recent years. Higher and higher dimensional systems can be prepared and controlled in optics experiments~\cite{stein,dada}. Therefore it is an important question to decide whether an experiment managed to create genuinely high-dimensional entanglement or the experimental data can alternatively be explained by assuming low dimensional entanglement. A measure which detects states with genuinely entangled degrees of freedom is based on the Schmidt number~\cite{schmidt_number}. Schmidt number $r$ of a bipartite system certifies that the state is entangled in at least $r$ degrees of freedom. 

As an illustration, let us consider the $d\times d$ isotropic state defined by
\begin{align}
\hat\rho_d^{\rm iso}(F)=F\ketbra{\Psi_d^+}{\Psi_d^+}+(1-F)\frac{\one-\ket{\Psi_d^+}\bra{\Psi_d^+}}{d^2-1},
\label{rhoiso}
\end{align} 
where 
\begin{align}
\ket{\Psi_d^+}=\frac{1}{\sqrt d}\sum_{k=1}^d\ket{k}\ket{k}
\label{maxent}
\end{align}
is the $d\times d$ maximally entangled state and $F$ is the entanglement fraction~\cite{hh} of the state $\hat\rho_d^{\rm iso}$. Based on the results of Ref.~\cite{schmidt_number} the state~(\ref{rhoiso}) can be shown to have Schmidt number at least $r$ for the entanglement fraction parameter
\begin{align}
F\ge \frac{r-1}{d},
\end{align}
where $r$ can take $r=(2,\ldots,d)$ in the above formula. Loosely speaking, $r$ is the minimum Schmidt rank of the pure states needed to construct it (later we will also give a formal definition of the Schmidt number). Hence tuning the parameter $F$ in a $d\times d$ isotropic state~(\ref{rhoiso}), we can change the Schmidt number $r$ of the state between two (i.e.~the case of standard entanglement) and $d$ (the maximum possible Schmidt number of a $d\times d$ state).

In this paper, our focus is on entangled states with a positive partial transpose (PPT). These are states which cannot be transformed into pure singlet states using local operations and classical communication. This procedure is called entanglement distillation~\cite{ED}. Such entangled states which cannot be distilled are called bound entangled~\cite{BE}, and they are not useful in protocols which are based on distillation. Nevertheless, PPT bound entangled states turn out to be useful in a couple of other quantum information tasks such as quantum key distribution~\cite{hllo,ozols}, superactivation of capacity of quantum channels~\cite{smithyard}, quantum metrology~\cite{Toth18}, EPR steering~\cite{tobias}, Bell-nonlocal correlations~\cite{VB,yuoh,PV17}, and channel discrimination~\cite{PW09}.    

Several interesting constructions of PPT states have been given in the literature (see e.g.~Refs.~\cite{BDM,breuer,PM,hhho,sentis,hllo}), however the question whether PPT states can be genuinely high-dimensionally entangled have been investigated only recently~\cite{szarek,chen,huber}. In particular, Huber et al.~\cite{huber} found that for a special class of $(d\times d)$-dimensional PPT entangled states the Schmidt number scales as $d/4$. This entails that one can generate PPT state with any number of genuinely entangled dimension provided the dimension $d$ is high enough. In the present work we strengthen this result by presenting a family of PPT states in dimensions $d\times d$ for which the Schmidt number scales as $d/2$. The proof relies on the special properties of the projections on the two-qudit symmetric and antisymmetric subspaces.

The structure of the paper is as follows. In Sec.~\ref{prelim} we give the necessary definitions including the definition of the Schmidt number and we define the class of PPT states investigated by Sindici and Piani~\cite{SinPia}. We recall the two methods of Ref.~\cite{SinPia} to prove that the PPT states in question are entangled, that is, their Schmidt number is greater than one. In Sec.~\ref{mainsec} we generalize the above methods of Sindici and Piani to lowerbound the Schmidt number of PPT states. In Sec.~\ref{firstres} we show how to construct PPT bound entangled states with Schmidt number $r\ge d/2$ for any even $d\ge 4$ dimension. In Sec.~\ref{techsec} a technical result is given concerning the normal form of an antisymmetric pure state. Then in Sec.~\ref{othermethod} we show that starting from an arbitrary entangled antisymmetric state $\hat\rho_A$, a semidefinite program~\cite{sdp} provides a lower bound on its Schmidt number. From this lower bound we in turn find PPT states with certain Schmidt number whose antisymmetric projections are proportional to $\hat\rho_A$. In Sec.~\ref{conc} we conclude our study and raise open questions.

\section{Preliminaries and overview of the two methods of Sindici and Piani}
\label{prelim} 

Any bipartite pure state can be written in the form of Schmidt decomposition:
\begin{equation}
|\varphi\rangle=\sum_{i=1}^{r}\sqrt{\pi_i}|a_i\rangle\otimes|b_i\rangle,
\label{eq:sch}
\end{equation}
where $\pi_i>0$, $\sum_i \pi_i=1$, $|a_i\rangle$ and $|b_i\rangle$ are orthonormal vectors in the component spaces and $r$, the Schmidt rank of $|\varphi\rangle$, is not larger than the dimensionality of either of the component spaces. A general state is represented by a positive semidefinite operator of trace one in the composite space, and it can be expressed as a convex combination of pure states (represented now by projections onto their state vectors) as:
\begin{equation}
\hat\rho=\sum_k q_k|\varphi_k\rangle\langle \varphi_k|.
\label{eq:mix}
\end{equation}
A mixed state can usually be written in many ways in forms like above. A state has Schmidt number $r$ if all possible such expressions contain at least one pure state whose Schmidt rank is at least $r$, and there is at least one expression in which neither of the Schmidt ranks exceeds $r$. From the definition it follows that for pure states the Schmidt number is the same as the Schmidt rank. A general state is separable if its Schmidt number is one, otherwise it is entangled.

In a recent work Sindici and Piani~\cite{SinPia} have given simple methods to construct entangled states with a positive partial transpose (PPT). The construction relied on properties of the antisymmetric and the symmetric subspaces of the joint Hilbert space of two systems of the same dimensionality. The projectors $\hat P_{\cal S}$ and $\hat P_{\cal A}$ defining the symmetric and the antisymmetric subspace, respectively, can be written as
\begin{align}
\hat P_{\cal S}=&\frac{\hat I+\hat V}{2}\label{eq:PS}\\
\hat P_{\cal A}=&\frac{\hat I-\hat V}{2},
\label{eq:PA}
\end{align}
where $\hat I$ and $\hat V$ are the identity and the swap operators, respectively. The swap operator can be written as
\begin{equation}
\hat V=\sum_{i=1}^d\sum_{j=1}^d|j,i\rangle\langle i,j|,
\label{eq:swapop}
\end{equation}
where $d$ is the dimensionality of the Hilbert spaces of the component systems. We use the shorthand notation of $|i,j\rangle\equiv |i\rangle\otimes|j\rangle$. The effect of $\hat V$ on a product state is $\hat V|\varphi_1\rangle\otimes|\varphi_2\rangle=|\varphi_2\rangle\otimes|\varphi_1\rangle$. The swap operator depends on the choice of local bases, states in the component spaces are regarded the same if their vector components are the same in the bases chosen. Therefore, the symmetric and the antisymmetric subspaces also depend on this choice. For example, any pure antisymmetric state becomes symmetric in the bases corresponding to its Schmidt decomposition. However, the symmetricity or antisymmetricity of states are preserved if both parties perform the same local transformation~\cite{werner}. In Ref.~\cite{SinPia} the authors consider only identical systems, where the choice of the bases is not completely arbitrary, but their methods are more general. As $\hat P_{\cal S}+\hat P_{\cal A}=\hat I$, the full space is the direct sum of the symmetric and antisymmetric subspaces. It is easy to check that the symmetric and the antisymmetric subspaces have dimensions $d_{\cal S}=d(d+1)/2$ and $d_{\cal A}=d(d-1)/2$, respectively.

The first method given in Ref.~\cite{SinPia} is based on semidefinite programming (SDP)~\cite{sdp}. One starts with an antisymmetric state $\hat\rho_{\cal A}$. Then the task is to find the PPT state $\hat\sigma$ (a positive semidefinite matrix of trace one, whose partial transpose is also positive semidefinite) whose projection to the antisymmetric subspace is proportional to the starting state (that is $\hat P_{\cal A}\hat\sigma\hat P_{\cal A}={\rm Tr}(\hat P_{\cal A}\hat\sigma)\hat\rho_{\cal A}$) and whose overlap ${\rm Tr}(\hat P_{\cal A}\hat\sigma)$ with the starting state is maximal. The procedure is successful if this maximal overlap denoted by $p^{PPT}(\hat\rho_{\cal A})$ is smaller than $1/2$, as it is proven in the paper~\cite{SinPia} that in this case the state $\hat\sigma$ is entangled. If the result is $1/2$, which is the upper bound for this quantity, one can start from another $\hat\rho_{\cal A}$. It has also been proven~\cite{SinPia} that if $\hat\sigma$ is an optimal solution of the problem above, so is $\hat P_{\cal A}\hat\sigma\hat P_{\cal A}+\hat P_{\cal S}\hat\sigma\hat P_{\cal S}=p^{PPT}(\hat\rho_{\cal A})\hat\rho_{\cal A}+(1- p^{PPT}(\hat\rho_{\cal A}))\hat\rho_{\cal S}^{opt}$, which is a convex mixture of the starting antisymmetric state and a symmetric state ($\hat P_{\cal S}\hat\sigma\hat P_{\cal S}$ properly normalized). Therefore, an alternative way to get $p^{PPT}(\hat\rho_{\cal A})$ is to find the optimal symmetric state whose convex mixture with $\hat\rho_{\cal A}$ is PPT, and the weight of $\hat\rho_{\cal A}$ in this mixed state is maximal. Then if this weight is less than $1/2$, the mixture is entangled.

The second method given in Ref.~\cite{SinPia} to construct a PPT entangled state is based on the fact that the Schmidt number of the projection of a separable state onto the antisymmetric subspace is always two. Therefore, any PPT state whose antisymmetric projection is proportional to a state whose Schmidt number is larger than two is necessarily entangled. For example, if we start from any antisymmetric state with a Schmidt number larger than two, and mix it with a symmetric state such that the mixture is PPT, the result will be a PPT entangled state. The symmetric state need not be the optimal one.

\section{Our results}
\label{mainsec}

\subsection{Construction of PPT states with any Schmidt number}
\label{firstres}

As we will show in this subsection, the second method of Sindici and Piani reviewed in Sec.~\ref{Intro} can be used to construct PPT bound entangled states with any Schmidt number. For that we will prove that the Schmidt number of the projection of a state onto the antisymmetric subspace $\hat P_{\cal A}\hat\sigma\hat P_{\cal A}$ can not be more than twice the Schmidt number of the state $\hat\sigma$. First let us consider a pure state $\ket{\varphi}$, whose Schmidt decomposition is defined by Eq.~(\ref{eq:sch}).
Then 
\begin{align}
\hat P_{\cal A}|\varphi\rangle=\frac{\sum_{i=1}^{r}\sqrt{\pi_i}(|a_i\rangle\otimes|b_i\rangle-|b_i\rangle\otimes|a_i\rangle)}{2},
\end{align}
where $r$ is the Schmidt rank of the state~(\ref{eq:sch}). Vectors $|a_i\rangle$ and $|b_i\rangle$ together can not span more than $2r$-dimensional subspaces of the component spaces, and the projected vector fully resides within the tensor product of those subspaces. Therefore, its Schmidt rank can not be more than $2r$. Now let us consider a mixed state. If it has Schmidt number $r$ it can be expressed as a convex mixture of pure states with Schmidt ranks at most $r$. If one applies the projection onto this form of the mixed state one gets an expansion in terms of vectors such that neither of them has Schmidt rank more than $2r$. Consequently, the Schmidt number of the projection can not be more than $2r$. Therefore, if one starts from an antisymmetric state whose Schmidt number is at least $r$ (where $r$ is even), and construct any PPT state whose antisymmetric projection is proportional to this state, then the Schmidt number of this PPT state has to be at least $r/2$. Again, we can do that by mixing the antisymmetric state with a sufficient amount of suitable symmetric state.

When choosing the antisymmetric state for the construction, it may be a problem that the Schmidt number of a mixed state is usually very hard to determine. However, as suggested in Ref.~\cite{SinPia}, one may start from a pure state. In case of $d$-dimensional component spaces, if $d$ is even, a generic antisymmetric pure state, which one can get by antisymmetrizing a random pure state has Schmidt rank $d$. By starting the construction from such a state with $d=2r$, one gets a PPT state with a Schmidt number at least $r$. In Ref.~\cite{SinPia} the authors considered the following states:
\begin{equation}
|\psi_{\cal A}\rangle=\sum_{\mu=1}^{d-1,{\rm odd}}c_{\mu}(|\mu,\mu+1\rangle-|\mu+1,\mu\rangle),
\label{eq:psia}
\end{equation}
where
\begin{equation}
\sum_{\mu=1}^{d-1,{\rm odd}}c^2_{\mu}=\frac{1}{2},\quad\quad c_{\mu}\geq 0
\label{eq:psianorm}
\end{equation}
If all $c_{\mu}>0$, the Schmidt rank of the state is $d$ (the Schmidt-coefficients are $c_{\mu}$, and each of them occurs twice). A special case is when all coefficients are equal:
\begin{equation}
|\psi_{0\cal A}\rangle=\frac{1}{\sqrt{d}}\sum_{\mu=1}^{d-1,{\rm odd}}(|\mu,\mu+1\rangle-|\mu+1,\mu\rangle).
\label{eq:psi0a}
\end{equation}
We will prove in the Appendix~\ref{app} that $p|\psi_{0\cal A}\rangle\langle\psi_{0\cal A}|+(1-p)\hat P_{\cal S}/d_{\cal S}$ with $p=1/(d+2)$ is a PPT bound entangled state with Schmidt number at least $d/2$, where $d$ is even.

\subsection{Normal form of antisymmetric pure states}
\label{techsec}

Now we will prove that any antisymmetric pure state may be transformed into the simple form given in Eq.~(\ref{eq:psia}) (for odd $d$ the summation goes from $1$ to $d-2$) with local transformations preserving the symmetric and antisymmetric subspaces. Let $|\varphi_{\cal A}\rangle$ be such an antisymmetric state in a $(d\times d)$-dimensional space: 
\begin{equation}
|\varphi_{\cal A}\rangle=\sum_{i=1}^{d}\sum_{j=1}^da_{ij}|i,j\rangle,
\label{eq:genant}
\end{equation}
where $A=(a_{ij})$ is a $d\times d$ skew symmetric matrix, that is $a_{ij}=-a_{ji}$. Then there exists a unitary matrix $U$~\cite{Zum62} such that
\begin{equation}
U^TAU=\left(\begin{array}{ccccc}
0&c_1&0&0&\dots\\
-c_1&0&0&0&\dots\\
0&0&0&c_3&\dots\\
0&0&-c_3&0&\dots\\
\vdots&\vdots&\vdots&\vdots&\ddots\\
\end{array}\right),
\label{eq:skewtr}
\end{equation}
where $c_{\mu}\geq 0$ ($\mu$ odd) are real numbers, and the last row and column contains zeros if $d$ is odd (a skew symmetric matrix of odd $d$ can not have full rank). The matrix on the right-hand side is called the normal form of $A$. The state characterized by such a matrix is just the one of Eq.~(\ref{eq:psia}), therefore we will also refer to it as the normal form. It is easy to see that this matrix transformation corresponds to both parties performing the same local basis transformation $|i'\rangle=\hat U^*|i\rangle$, where $\hat U^*=\sum_{kl}U^*_{kl}|k\rangle\langle l|$. Such simultaneous local transformations preserve the antisymmetricity and the symmetricity of a state ($\hat U^*\otimes\hat U^*$ commutes with both $P_{\cal A}$ and $P_{\cal S}$). Starting from $\hat\rho_{\cal A}$, the constructions considered above will give the same results in the transformed local basis as in the original one. Therefore, we may calculate $p^{PPT}$ for any pure antisymmetric state after transforming it into its normal form. From the normal form it is also clear that all antisymmetric pure states have an even Schmidt rank in any number of dimensions, and from this it follows that all antisymmetric mixed states have an even Schmidt number. We will also use later that if the Schmidt rank $r$ of a $(d\times d)$-dimensional pure state in its normal form is less than $d$, then the subspace necessary to accommodate the state is spanned by $r$ basis vectors in each component space. 

\subsection{Schmidt number from the value of $p^{PPT}(\hat\rho_{\cal A})$}
\label{othermethod}

The important result of the first method of Ref.~\cite{SinPia} is based on the observation that $p^{PPT}(\hat\rho_{\cal A})<1/2$ proves that the PPT state resulting from the construction is entangled. We will show that from the value of $p^{PPT}(\hat\rho_{\cal A})$ we may be able to tell that the Schmidt number of $\hat\rho_{\cal A}$ is larger than a certain value, consequently any PPT state constructed from it with the methods given above has Schmidt number at least half that value. It has been proven~\cite{SinPia} that there exists a lower bound for $p^{PPT}(\hat\rho_{\cal A})$ depending on the dimension of the component spaces $d$, which decreases monotonously as a function of $d$. The bound derived is $2/[d(d+1)+2]$, which is not tight, as we will show it later by deriving a much better one. Let us denote the $d$-dependent monotonously decreasing lower bound by $L(d)$. What we will show is that $L(d)$ is actually $L(r)$, where $r$ is the Schmidt number of $\hat\rho_{\cal A}$, that is in any space dimensions $p^{PPT}(\hat\rho_{\cal A})$ can not be smaller than $L(r)$. First let us consider a pure state in a $d\times d$-dimensional space with Schmidt rank $r<d$. We are allowed to transform it into its normal form. Let us reduce the space to the $r\times r$-dimensional subspace that accommodates the state by dropping the basis vectors orthogonal to this subspace. Determined in this space $p^{PPT}$ can not be smaller than $L(r)$. If the dropped basis vectors are added back, $p^{PPT}$ can not decrease, because it is the maximum of the overlaps between the state and all PPT states $\sigma$ whose antisymmetric projection is proportional to the state, and the extended space means an even larger variety of appropriate $\sigma$ states. Now let us consider mixed states. First we will show that $p^{PPT}$ belonging to a convex mixture of two antisymmetric states can not be smaller than $p^{PPT}$ belonging to any of the two states. Let $p_1\equiv p^{PPT}(\hat\rho_{1\cal A})$, $p_2\equiv p^{PPT}(\hat\rho_{2\cal A})$ and $p_1<p_2$. Then there exist symmetric states $\hat\rho_{1\cal S}$ and $\hat\rho_{2\cal S}$ such that $p_1\hat\rho_{1\cal A}+(1-p_1)\hat\rho_{1\cal S}$ and $p_2\hat\rho_{2\cal A}+(1-p_2)\hat\rho_{2\cal S}$ are PPT. Let us mix the second state with any symmetric PPT state such that the weight of $\hat\rho_{2\cal A}$ in the mixture is reduced to $p_1$. We get  $p_1\hat\rho_{2\cal A}+(1-p_1)\hat\rho'_{2\cal S}$, which is also a PPT state. Then $p_1[\lambda\hat\rho_{1\cal A}+(1-\lambda)\hat\rho_{2\cal A}]+(1-p_1)[\lambda\hat\rho_{1\cal S}+(1-\lambda)\hat\rho_{2\cal S}]$, where $0\leq\lambda\leq 1$ is also PPT, which proves that $p^{PPT}$ belonging to the convex mixture is not smaller than $p_1$. Then it follows that $p^{PPT}$ belonging to a convex mixture of any number of states is not smaller than the smallest of $p^{PPT}$ belonging to the components: mixing can not make $p^{PPT}$ smaller. A mixed state of Schmidt number $r$ can be written as a convex mixture of pure states with a maximum Schmidt number of $r$. As $p^{PPT}$ of any of the components can not be smaller than the lower bound belonging to $r$, $p^{PPT}$ belonging to the mixture can not be smaller either, which proves the statement for mixed states as well. The result means that if $p^{PPT}(\hat\rho_{\cal A})<L(d)$, its Schmidt number is larger than $d$ (actually, due to the non-existence of antisymmetric states with odd Schmidt numbers, it has to be at least $d+2$).

In the Appendix~\ref{app} we will show that $L(d)=1/(d+2)$ for even $d\geq 4$ and $L(2)=1/2$ are lower bounds for $p^{PPT}(\hat\rho_{\cal A})$. For $d=2$ any smaller value would imply the existence of a bound entangled state. For odd $d$ the bound has to be the same as for $d-1$. There are no antisymmetric states with an odd Schmidt number, so the maximum Schmidt number of $\hat\rho_{\cal A}$ is $d-1$. However, as we have shown earlier, $p^{PPT}(\hat\rho_{\cal A})$ can not be smaller than the lower bound corresponding to its Schmidt number. Therefore, $L(3)=1/2$, and for odd $d\geq 5$ it is $L(d)=1/(d+1)$.

To prove the lower bound for even $d$ we have given above, it is enough to consider pure states, because mixing can not make $p^{PPT}$ smaller, as we have shown earlier. We may also confine ourselves to their normal form given by Eq.~\ref{eq:psia}.

The first step of the proof given in the Appendix~\ref{app} is to take the state $|\psi_{0\cal A}\rangle$ with equal amplitudes given by Eq.~(\ref{eq:psi0a}), and calculate how much admixture of $\hat P_{\cal S}/d_{\cal S}$ is necessary to make it a PPT state. This is the same symmetric state considered in Theorem 1 of Ref.~\cite{SinPia} calculating the lower bound for any state. Their bound is not tight because for the antisymmetric state for which this choice is optimal less admixture is enough (that is $p^{PPT}$ is larger), while for all other states the choice is not optimal. From numerical calculations in smaller dimensions we believe that for $|\psi_{0\cal A}\rangle$ this choice is actually optimal, therefore the lower bound $1/(d+2)$ we get is the true value of $p^{PPT}(|\psi_{0\cal A}\rangle\langle\psi_{0\cal A}|)$. Then we show that for states of normal form with non-equal amplitudes $p^{PPT}$ is larger, therefore, $1/(d+2)$ is a lower bound for them, too. Our conjecture is that this bound is tight.

As $L(2)=1/2$, any smaller value for $p^{PPT}(\hat\rho_{\cal A})$ proves that the Schmidt number of $\hat\rho_{\cal A}$ is at least $4$, which also means that any PPT state whose antisymmetric projection is proportional to $\hat\rho_{\cal A}$ is entangled (that is its Schmidt number is at least 2). This is true for any such PPT state, not only for the one that comes out of the construction providing the value of $p^{PPT}(\hat\rho_{\cal A})$. For $d\geq 4$, $p^{PPT}(\hat\rho_{\cal A})<L(d)$ with $L(d)=1/(d+2)$ means that the Schmidt number of $\hat\rho_{\cal A}$ is at least $d+2$, and all PPT states whose antisymmetric projections are proportional to $\hat\rho_{\cal A}$ have Schmidt numbers at least $(d/2)+1$.

\section{Conclusions}
\label{conc}
In this paper, we proved that a class of $d\times d$ states with a positive partial transposition gives rise to Schmidt number $d/2$. This family is the one investigated by Sindici and Piani~\cite{SinPia} with respect to their entanglement properties. Here we generalized their methods to explore the dimensionality of the entanglement of such states. In particular, the Schmidt number $d/2$ of these states improves the Schmidt number of the states investigated by Huber et al.~\cite{huber} which latter scales as $d/4$. 

There are a couple of open questions left. We first ask whether such highly entangled states are useful for communication tasks beyond known protocols. Secondly, is it possible to construct PPT states, which have Schmidt number higher than $d/2$? In particular, is it possible to approach the value of $(d-1)$? In the recent paper~\cite{YLT}, it has been shown that no ($3\times 3$)-dimensional PPT state has Schmidt number three. In this respect, one may raise the question: what is the dimensionality of the smallest PPT state which has Schmidt number at least three? Our present result provides such a PPT state for dimension $d=6$. But is there possibly a smaller dimensional example? Finally, we believe that our investigations are relevant from an experimental point of view as well given the recent advances in the experimental implementation of bound entangled states~\cite{exp1,exp2,exp3}.

\section{Acknowledgements} This work was supported by the National Research, Development and Innovation Office NKFIH (Grant No. KH125096).




\appendix\section{Lower bound for $\boldsymbol{p^{PPT}(\hat\rho_{\cal A})}$}
\label{app}

As we have shown in the main text, to calculate the $d$-dependent lower bound $L(d)$ for $p^{PPT}(\hat\rho_{\cal A})$ it is enough to consider pure states in their normal form in $(d\times d)$-dimensional spaces, where $d$ is even.

First we determine a lower bound of $p^{PPT}$ for the special state $|\psi_{0\cal A}\rangle$ given by Eq.~(\ref{eq:psi0a}). The problem we are going to solve is to find the largest value of $p$ such that the partial transpose of
\begin{equation}
\hat\sigma_0=p|\psi_{0\cal A}\rangle\langle\psi_{0\cal A}|+(1-p)\frac{\hat P_{\cal S}}{d_{\cal S}}.
\label{eq:rho0}
\end{equation}
is positive semidefinite. In particular, we show that this value is $1/(d+2)$. If $\hat P_{\cal S}/d_{\cal S}$ is the optimal symmetric operator, then this way we get $p^{PPT}(|\psi_{0{\cal A}}\rangle\langle\psi_{0{\cal A}}|)$ itself, if not, we get a lower bound. For smaller $d$ values we have numerical evidence that this is the optimal choice, and our conjecture is that this is true for any dimensions. The partial transposes (denoted by upper index $\Gamma$) of the operators appearing in the right-hand side of Eq.~(\ref{eq:rho0}) are:
\begin{align}
&|\psi_{0\cal A}\rangle\langle\psi_{0\cal A}|^\Gamma=\frac{1}{d}\sum_{\mu=1}^{d-1,{\rm odd}}\sum_{\nu=1}^{d-1,{\rm odd}}\nonumber\\
&(|\mu,\nu+1\rangle\langle\nu,\mu+1|+|\mu+1,\nu\rangle\langle\nu+1,\mu|-\nonumber\\
&|\mu,\nu\rangle\langle\nu+1,\mu+1|-|\nu+1,\mu+1)\rangle\langle\mu,\nu|),
\label{eq:pasimPT}
\end{align}
and
\begin{align}
&\frac{\hat P^{\Gamma}_{\cal S}}{d_{\cal S}}=\frac{\sum_{i=1}^d\sum_{j=1}^d|i,j\rangle\langle i,j|+\sum_{i=1}^d\sum_{j=1}^d|i,i\rangle\langle j,j|}{d(d+1)}.
\label{eq:psimPT}
\end{align}
Eq.~(\ref{eq:pasimPT}) can be derived from Eq.~(\ref{eq:psi0a}), while Eq.~(\ref{eq:psimPT}) from Eqs.~(\ref{eq:PS}) and (\ref{eq:swapop}), and $d_{\cal S}=d(d+1)/2$.

Now we will solve the eigenvalue problem, and determine the largest $p$ value such that none of the eigenvalues are negative. The calculation can be simplified by realizing that the matrix representing operator $\hat\sigma^{\Gamma}_0$ has a block-diagonal structure. This structure is not apparent, as the rows and columns belonging to each block do not follow each other. Nevertheless, if there exists a subset of indices such that the rows and columns corresponding to those indices have nonzero elements only where they intersect, then these rows and columns belong to a block of the matrix. Such blocks can be treated separately when solving the eigenvalue problem.

The first double sum of the symmetric part of the operator, which is given by Eq.~(\ref{eq:psimPT}), is diagonal (proportional to the identity operator), while the second one is a $d\times d$ block with equal matrix elements.

The pair of positive terms of Eq.~(\ref{eq:pasimPT}), $(1/d)|\mu,\nu+1\rangle\langle\nu,\mu+1|$ and 
$(1/d)|\nu,\mu+1\rangle\langle\mu,\nu+1|$, and similarly the $(1/d)|\mu+1,\nu\rangle\langle\nu+1,\mu|$ and $(1/d)|\nu+1,\mu\rangle\langle\mu+1,\nu|$ correspond to off-diagonal elements of $2\times 2$ blocks if $\nu\neq\mu$ (they exist if $d\geq 4$). Each of these blocks of $\hat\sigma^{\Gamma}_0$, including the diagonal elements coming from the symmetric matrix given by Eq.~(\ref{eq:psimPT}) look like:
\begin{equation}
\left(\begin{array}{cc}
\frac{1-p}{d(d+1)}&\frac{p}{d}\\
\frac{p}{d}&\frac{1-p}{d(d+1)}\\
\end{array}\right).
\label{eq:2X2}
\end{equation}
The solutions of the eigenvalue equation of this block are $(1+pd)/[d(d+1)]$ , which is always positive, and $[1-(d+2)p]/[d(d+1)]$, which is non-negative if $p\leq 1/(d+2)$. There are $d(d-2)/4$ such blocks. Each of the positive terms of Eq.~(\ref{eq:pasimPT}) with $\nu=\mu$ are diagonal elements. Together with the contribution coming from Eq.~(\ref{eq:psimPT}) each of them gives the always positive eigenvalue of $(1+pd)/[d(d+1)]$ again. There are $d$ such diagonal elements.
 
The pair of negative terms of Eq.~(\ref{eq:pasimPT}): $-(1/d)|\mu,\nu\rangle\langle\nu+1,\mu+1|$ and $-(1/d)|\nu+1,\mu+1\rangle\langle\mu,\nu|$ also correspond to off-diagonal elements of $2\times 2$ blocks. If $\nu=\mu$ ($d/2$ pairs) they fall within the $d\times d$ block of the matrix corresponding to Eq.~(\ref{eq:psimPT}) whose eigenvalue problem we will solve in the next step. The $\nu\neq\mu$ cases ($d(d-2)/4$ pairs) lead to the same $2\times 2$ matrix as given in Eq.~(\ref{eq:2X2}) but with negative signs for the off-diagonal elements, giving the same two eigenvalues as we have already got. 

Now we will calculate the eigenvalues of the $d\times d$ block. We will demonstrate the calculations on the example of $d=8$. The determinant to be calculated is the following: 
\begin{equation}
D_8=\left|\begin{array}{cccccccc}
a&c&b&b&b&b&b&b\\
c&a&b&b&b&b&b&b\\
b&b&a&c&b&b&b&b\\
b&b&c&a&b&b&b&b\\
b&b&b&b&a&c&b&b\\
b&b&b&b&c&a&b&b\\
b&b&b&b&b&b&a&c\\
b&b&b&b&b&b&c&a\\
\end{array}\right|,
\label{eq:D8or}
\end{equation}
where
\begin{align}
a&=2\frac{1-p}{d(d+1)}-\lambda,\nonumber\\
b&=\frac{1-p}{d(d+1)},\nonumber\\
c&=b-\frac{p}{d}=\frac{1-p(d+2)}{d(d+1)}.
\label{eq:D8notat}
\end{align}
The first term of $a$ and $b$ comes from Eq.~(\ref{eq:psimPT}), while the $-p/d$ term of $c$ comes from Eq.~(\ref{eq:pasimPT}), and $\lambda$ is the eigenvalue. Let us subtract the $\nu$th row from the $(\nu-1)$th one then the $\nu$th column from the $(\nu-1)$th one for all odd $\nu$, and introduce the notation $f\equiv a-c$ to arrive at:
\begin{equation}
D_8=
\left|\begin{array}{cccccccc}
a&-f&b&0&b&0&b&0\\
-f&2f&0&0&0&0&0&0\\
b&0&a&-f&b&0&b&0\\
0&0&-f&2f&0&0&0&0\\
b&0&b&0&a&-f&b&0\\
0&0&0&0&-f&2f&0&0\\
b&0&b&0&b&0&a&-f\\
0&0&0&0&0&0&-f&2f\\
\end{array}\right|.
\label{eq:D8s1}
\end{equation}
Let us subtract the 5th row from the 7th one, then the 3th row from the 5th one and finally the first row from the third one. Then let us do the same with the corresponding columns. Introducing the notation $e\equiv a-b$ we get:
\begin{equation}
D_8=
\left|\begin{array}{cccccccc}
a&-f&-e&0&0&0&0&0\\
-f&2f&f&0&0&0&0&0\\
-e&f&2e&-f&-e&0&0&0\\
0&0&-f&2f&f&0&0&0\\
0&0&-e&f&2e&-f&-e&0\\
0&0&0&0&-f&2f&f&0\\
0&0&0&0&-e&f&2e&-f\\
0&0&0&0&0&0&-f&2f\\
\end{array}\right|.
\label{eq:D8s2}
\end{equation}
Finally, let us add one half times the 8th column to the 7th one, subtract one half times the 6th row from the 7th one, and then subtract one half times the 6th column from the 7th one to arrive at:
\begin{equation}
D_8=
\left|\begin{array}{cccccccc}
a&-f&-e&0&0&0&0&0\\
-f&2f&f&0&0&0&0&0\\
-e&f&2e&-f&-e&0&0&0\\
0&0&-f&2f&f&0&0&0\\
0&0&-e&f&2e&-f&-e+\frac{f}{2}&0\\
0&0&0&0&-f&2f&0&0\\
0&0&0&0&-e+\frac{f}{2}&0&2e-f&-f\\
0&0&0&0&0&0&0&2f\\
\end{array}\right|.
\label{eq:D8s3}
\end{equation}
It is easy to check that the manipulations we have done have not changed the values of the subdeterminants $D_6$, $D_4$ and $D_2$. Then it is not difficult to see that $D_8$ can be written as:
\begin{equation}
D_8=2f\left[(2e-f)D_6-\left(e-\frac{f}{2}\right)^22fD_4\right].
\label{eq:recur8}
\end{equation}
We could have arrived at an analogous recurrence relation for any even $d\geq 6$. Using $e\equiv a-b$ and $f\equiv a-c$ the general recurrence relation is:
\begin{equation}
D_{d+4}=(a-c)(a-2b+c)[2D_{d+2}-(a-c)(a-2b+c)D_{d}].
\label{eq:recurg}
\end{equation}
The solution of the relation with the correct initial values is:
\begin{equation}
D_{d}=(a-c)^{\frac{d}{2}}(a-2b+c)^{\frac{d}{2}-1}[a+(d-2)b+c].
\label{eq:detval}
\end{equation}
This can be proven by induction: $D_2$ and $D_4$ can explicitly be calculated and compared, and it is easy to check that the expression satisfies the recurrence relation. Using Eq.~(\ref{eq:D8notat}) the factors appearing in Eq.~(\ref{eq:detval}) are the following:
\begin{align}
a-c&=\frac{1+pd}{d(d+1)}-\lambda,\\
a-2b+c&=\frac{1-p(d+2)}{d(d+1)}-\lambda,\\
a+(d-2)b+c&=\frac{1+d-p(d+4)}{d(d+1)}-\lambda.
\label{eq:detfactors}
\end{align}
A root of $(1+pd)/[d(d+1)]$ with multiplicity $d/2$ of the eigenvalue equation $D_d=0$ comes from $a-c=0$. This root is the always positive solution we have already got. If $d\geq 4$ we get root $[1-(d+2)p]/[d(d+1)]$ with multiplicity $(d/2-1)$ from requiring $a-2b+c=0$. We have already got this root as well from the $2\times 2$ blocks. The remaining root, $[1+d-p(d+4)]/[d(d+1)]$ comes from $a+(d-2)b+c=0$.

To summarize the results above, for $d\geq 4$ there are three different eigenvalues of operator $\hat\sigma^{\Gamma}_0$. The always positive $(1+pd)/[d(d+1)]$ has a multiplicity of $d(d+1)/2$, coming from the $d\times d$ block, the $2\times 2$ blocks and the single diagonal elements. Eigenvalue $[1-(d+2)p]/[d(d+1)]$ has multiplicity $(d+1)(d-2)/2$ coming from the $d\times d$ block and the $2\times 2$ blocks. This is non-negative if $p\leq 1/(d+2)$. The last root $[1+d-p(d+4)]/[d(d+1)]$ is a single one, and it is non-negative if $p\leq (d+1)/(d+4)$, a less strict condition than the previous one. Therefore, for $d\geq 4$ the largest value of $p$ such that $\hat\sigma^{\Gamma}_0$ is positive semidefinite is $p=1/(d+2)$. For $d=2$ the eigenvalue giving this condition does not exist, therefore we get the appropriate condition $p=0.5$ from the last eigenvalue.

The last step is to prove that $p^{PPT}(|\psi_{{\cal A}}\rangle\langle\psi_{{\cal A}}|)\geq p^{PPT}(|\psi_{0{\cal A}}\rangle\langle\psi_{0{\cal A}}|)$. Let us define operator $\hat\tau$ as:
\begin{equation}
\hat\tau\equiv\sum_{i=1}^d\sum_{j=1}^d t_i t_j|i,j\rangle\langle i,j|,
\label{eq:tauoper}
\end{equation}
where
\begin{equation}
t_{\mu}=t_{\mu+1}\equiv\sqrt{c_{\mu}}d^{1/4}\quad\quad\mu\hphantom{0}{\rm odd},
\label{eq:tfac}
\end{equation}
and $c_{\mu}$ are the coefficients appearing in Eq.~(\ref{eq:psia}), the definition of $|\psi_{{\cal A}}\rangle$. From Eqs.~(\ref{eq:psia})-(\ref{eq:psi0a}) it is easy to see that
\begin{align}
|\psi_{\cal A}\rangle&=\hat\tau|\psi_{0\cal A}\rangle,\\
\sum_{i=1}^d t_i^4&=d.
\label{eq:tauprop}
\end{align}
The matrix of operator $\hat\tau$ is diagonal with non-negative entries. If operator $\hat R$ is positive semidefinite, so is $\hat\tau\hat R\hat\tau$, because if the expectation value of the latter with a state $|\xi\rangle$ were negative, so were the expectation value of the former with $\hat\tau|\xi\rangle$. Furthermore, it commutes with the partial transposition, as it multiplies both $|i,j\rangle\langle k,l|$ and $|i,l\rangle\langle k,j|$ with the same factor of $t_it_jt_kt_l$. It also preserves the symmetricity or antisymmetricity of a state. Let us take $\hat\tau\hat\sigma_0\hat\tau$ (see Eq.~(\ref{eq:rho0})), that is:
\begin{equation}
p|\psi_{\cal A}\rangle\langle\psi_{\cal A}|+(1-p)\frac{\hat\tau\hat P_{\cal S}\hat\tau}{d_{\cal S}}
\label{eq:notrho}
\end{equation}
with $p=1/(d+2)$, which is the lower bound for $p^{PPT}(|\psi_{0\cal A}\rangle\langle\psi_{0\cal A}|)$. Due to the properties of operator $\hat\tau$ stated above, the partial transpose of the operator is positive semidefinite with this value of $p$. The second term is a symmetric positive semidefinite operator. However, the operator in Eq.~(\ref{eq:notrho}) is not a density operator, because $\hat\tau\hat P_{\cal S}\hat\tau/d_{\cal S}$ in the second term is not one.  Projector $\hat P_{\cal S}$ is an equal combination of the swap and the identity operators (see Eq.~(\ref{eq:PS})). For the transformed swap operator $Tr(\hat\tau\sum_{ij}|i,j\rangle\langle j,i|\hat\tau)=\sum_i t_i^4=d$, which is the same as for the swap operator itself. However, for the transformed identity operator the trace is $\sum_{ij}t_i^2t_j^2=(\sum_i t_i^2)^2\leq d\sum_i t_i^4=d^2$, that is if $t_i$ are not all equal, it is smaller than the trace of the identity operator. Therefore, to make the operator in Eq.~(\ref{eq:notrho}) a density operator that is a proper mixture of the antisymmetric operator and a symmetric one to provide a lower bound for $p^{PPT}$, the second term has to be renormalized by multiplying it by a factor larger than one. This renormalization will not only preserve the PPT property of the operator with the same $p$, but it will even allow $p$ to be increased somewhat and still having a PPT operator. Therefore, $p^{PPT}(|\psi_{{\cal A}}\rangle\langle\psi_{{\cal A}}|)\geq p^{PPT}(|\psi_{0{\cal A}}\rangle\langle\psi_{0{\cal A}}|)$, indeed.
 
\end{document}